\documentclass{moriond}

\usepackage{amsmath}
\usepackage{xcolor} % in preamble

\def\be{\begin{equation}}
\def\ee{\end{equation}}
\def\bea{\begin{eqnarray}}
\def\eea{\end{eqnarray}}

\newcommand{\Photo}{}

\begin{document}
\vspace*{4cm}
\title{PHENIX Measurements of Light Hadron and Vector Meson Production at RHIC}

\author{ MURAD SARSOUR (on behalf of the PHENIX Collaboration) }

\address{Department of Physics and Astronomy, Georgia State University, \\
Atlanta, GA 30303, USA}

\maketitle

\abstracts{
Measurements of light hadron production in ultrarelativistic nuclear collisions provide essential insight into final-state effects arising from both hot and cold nuclear matter. They probe collective behavior, hadronization via recombination, and baryon and strangeness enhancement, while their system-size and centrality dependence constrain the role of initial-state geometry and nuclear parton distribution functions. In this talk, we present recent PHENIX measurements of identified charged hadrons ($\pi/K/p$) at midrapidity ($|y| < 0.35$) and low-mass vector mesons, including $\omega$, $\rho$, and $\phi$, at forward rapidity ($1.2 < |y| < 2.2$) in $p+p$, $p+$Al, $p/d/^{3}$He+Cu+Au, and Au+Au collisions at $\sqrt{s_{NN}} = 200$~GeV, as well as U+U collisions at $\sqrt{s_{NN}} = 193$~GeV. Tests of various empirical scaling behaviors, together with comparisons to previous measurements and theoretical model calculations, are discussed.
}

\section{Introduction}

Understanding the dynamics of strongly interacting matter in high-energy nuclear collisions is a central goal of quantum chromodynamics (QCD). Experiments at RHIC and the LHC have established the formation of a deconfined quark--gluon plasma (QGP), which exhibits collective behavior consistent with nearly perfect fluidity~\cite{Adcox:2004mh,Heinz:2013th}. A key open question is how final-state observables are influenced by the interplay between system size, collision geometry, and microscopic particle production mechanisms.

Identified hadron measurements provide a powerful tool to address these questions. By studying particle yields and spectra across different collision systems, one can test whether particle production follows universal scaling with global event properties such as multiplicity, or whether more differential features play a dominant role~\cite{Adam:2016dau,Andronic:2017pug}. Comparisons across hadron species further probe the underlying dynamics of hadronization, with species-dependent effects often interpreted in terms of collective flow and quark recombination~\cite{Fries:2003vb}. In addition, measurements at forward rapidity provide sensitivity to different kinematic regimes and partonic momentum fractions, enabling the study of flavor-dependent effects beyond midrapidity and offering new constraints on particle production mechanisms.

Complementary observables extend this picture. In particular, correlations involving $J/\psi$ production provide an additional handle on the connection between soft and hard processes, as well as on the role of the medium in shaping heavy-quark dynamics. Such measurements are sensitive to both initial- and final-state effects, and establish a link between bulk particle production and quarkonium observables.

In this contribution, we present recent PHENIX measurements of identified particle production across multiple collision systems, with an emphasis on new forward-rapidity vector meson results and first studies of $J/\psi$ production as a function of event multiplicity with controlled rapidity gaps. These measurements provide new constraints on system-size scaling, species-dependent effects, and the interplay between soft and hard processes.

The measurements were performed with the PHENIX detector at RHIC, a multipurpose experiment optimized for the detection of leptons, photons, and hadrons. It comprises central arm spectrometers at midrapidity ($|y| < 0.35$) and forward muon arms covering $1.2 < |y| < 2.4$, enabling complementary measurements across a broad rapidity range~\cite{ADCOX2003469}.

\section{Results}

\begin{figure}[h]
\centering
\colorbox{white}{\includegraphics[width=\linewidth]{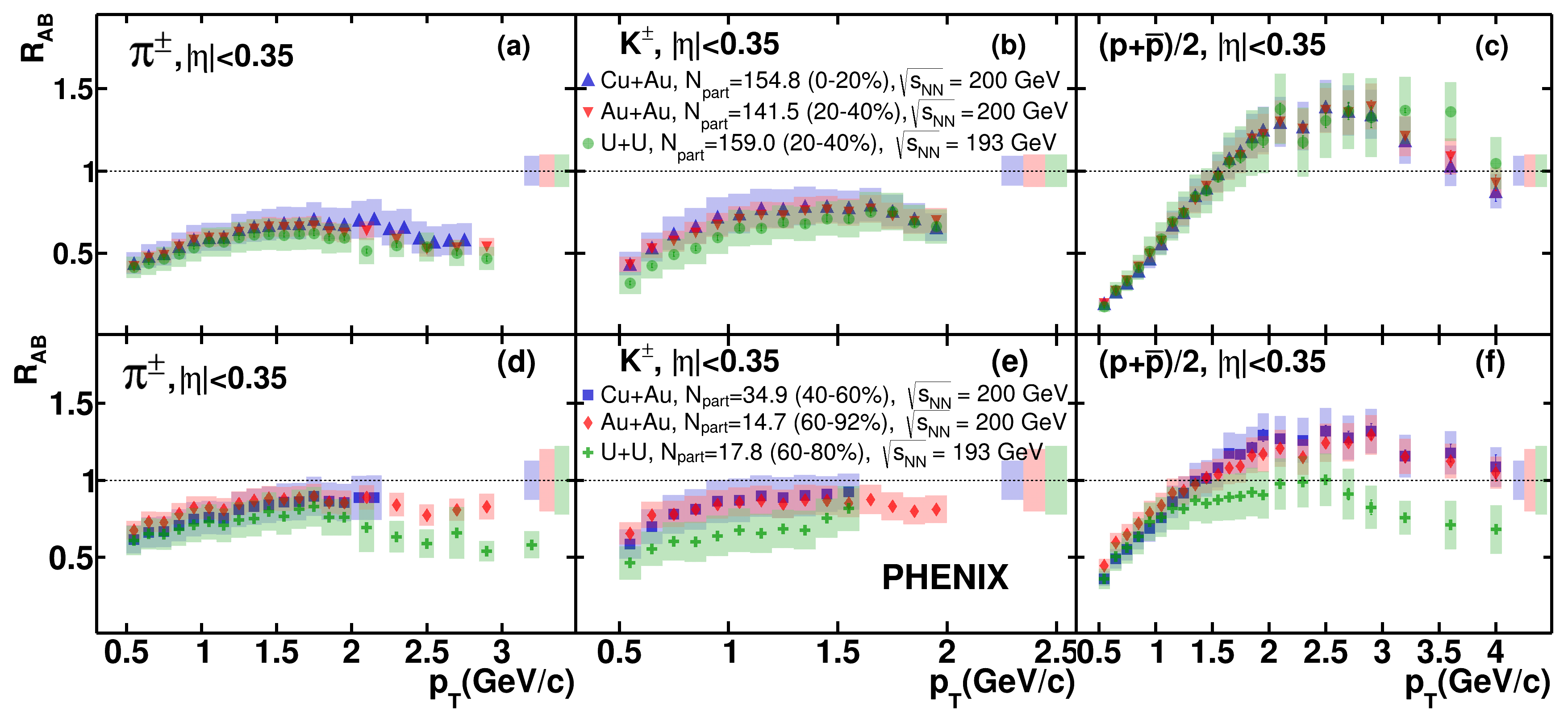}}
\caption{$R_{AB}$ as a function of $p_T$ measured in central and peripheral Cu + Au, Au + Au, and U + U collisions.
}
\label{fig:system_scaling}
\end{figure}

The investigation begins with the global behavior of particle production as a function of the number of participating nucleons ($N_{part}$). Measurements of the nuclear modification factor  $R_{AB}$ for $\pi^{\pm}$, $K^{\pm}$, and $(p+\overline{p})/2$ in Cu+Au, Au+Au, and U+U systems at $\sqrt{s_{NN}} \approx 200$~GeV are compared in Fig.~\ref{fig:system_scaling} ~\cite{PRC109_2024}.
Across these diverse systems, $R_{AB}$ values for each species are found to be in close agreement when compared at similar $N_{part}$, despite differences in the initial-state geometry. This suggests that identified hadron production is largely controlled by system size, with residual sensitivity to geometry not excluded.
In central collisions (top panels), mesons exhibit strong suppression at intermediate $p_{T}$, consistent with significant partonic energy loss in a dense medium. Conversely, a persistent baryon-meson splitting is observed where protons are less suppressed than pions and kaons. This feature, which diminishes as the systems become more peripheral (bottom panels), is commonly interpreted as evidence for hadronization via parton recombination in a strongly interacting medium~\cite{Fries:2003vb}. Overall, these results support a picture in which final-state effects governed by the bulk properties of the quark-gluon plasma drive both the suppression patterns and the resulting hadron composition~\cite{PRC109_2024}.

\begin{figure}[h]
\centering
\includegraphics[width=0.5\linewidth]{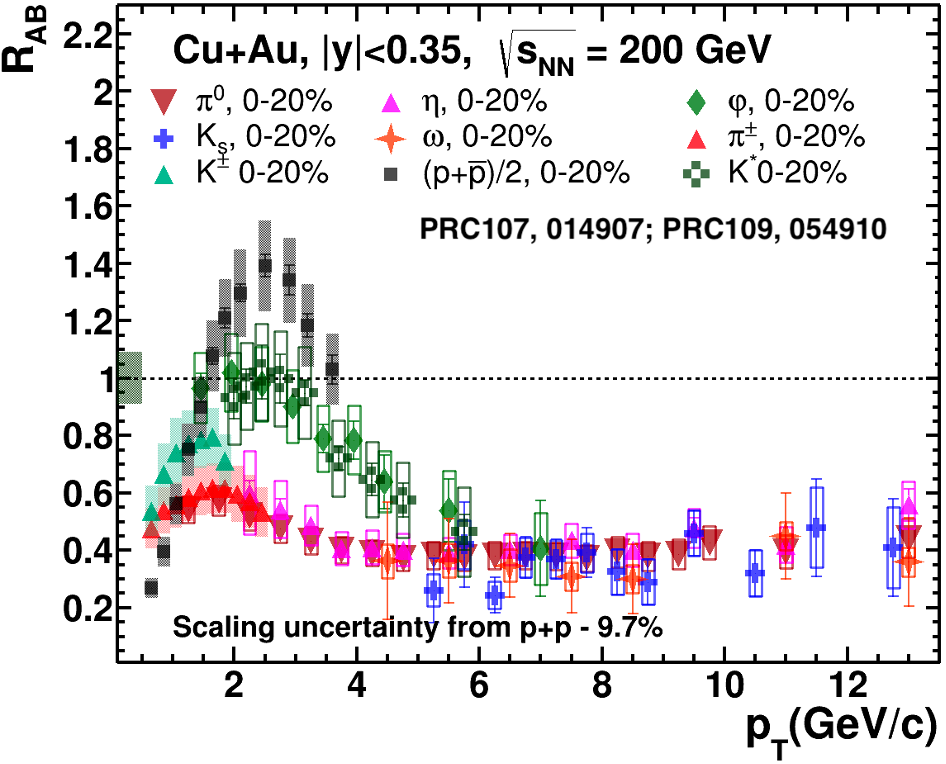}
\caption{Summary of $p_T$ dependence by particle species in Cu+Au collisions.}
\label{fig:species_dep}
\end{figure}
The transverse momentum dependence of identified hadrons is examined in Fig.~\ref{fig:species_dep}, which summarizes the species dependence of $R_{AB}$ across a wide $p_T$ range~\cite{PRC109_2024,PhysRevC.107.014907}.
The $R_{AB}$ for various species ($\pi^0, \eta, \phi, \omega, \pi^{\pm}, K^{\pm}, p+\overline{p}$) is shown to converge at high transverse momentum ($p_T > 6~\text{GeV}/c$). This universal suppression indicates that high-$p_T$ hadron production is dominated by partonic energy loss, with limited sensitivity to hadron species.
In contrast, the intermediate $p_T$ region ($2 < p_T < 5~\text{GeV}/c$) in Fig.~\ref{fig:species_dep} exhibits significant flavor dependence. The distinct separation between baryons (protons, which peak above unity) and mesons (pions and kaons, which remain suppressed) is a strong indicator of hadronization via quark recombination.
This behavior supports a picture in which hadronization at intermediate $p_T$ is influenced by quark recombination, while high-$p_T$ production is dominated by partonic energy loss.

\begin{figure}[h]
\centering
\includegraphics[width=0.45\linewidth]{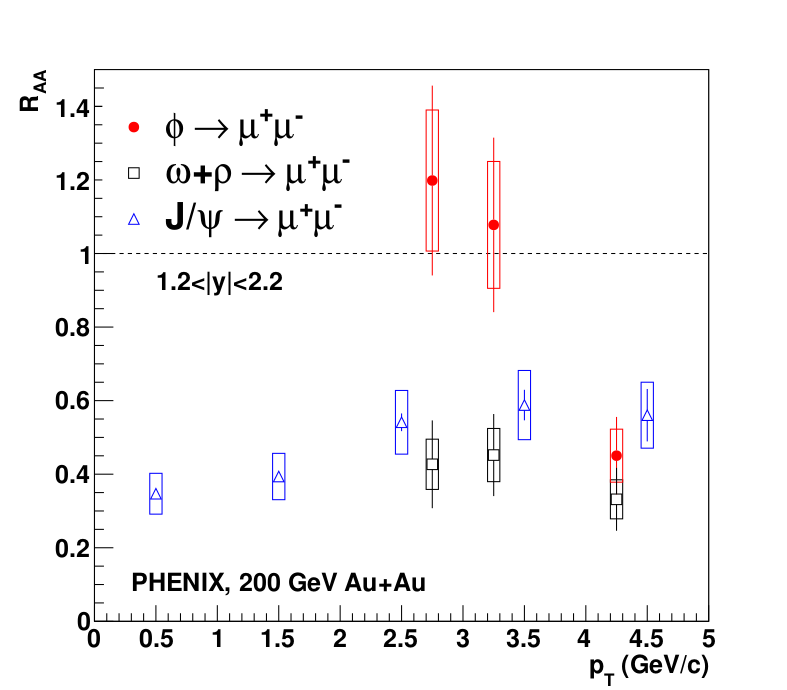}
\includegraphics[width=0.45\linewidth]{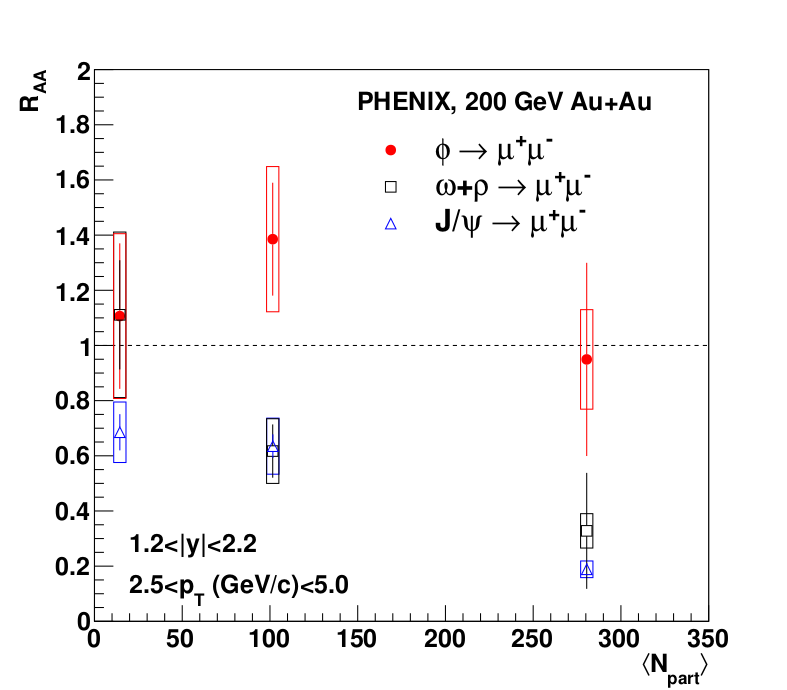}
\caption{$R_{AA}$ as a function of $p_T$ (left) and $\langle N_{part} \rangle$ (right), demonstrating species-independent suppression at high momentum and large system.}
\label{fig:flavor_dep}
\end{figure}
Figure~\ref{fig:flavor_dep} displays the $R_{AA}$ for vector mesons at forward rapidity ($1.2 < |y| < 2.2$) in $Au+Au$ collisions~\cite{PRC112_2025}. The left panel shows that while the $\omega+\rho$ sum (open squares) and $J/\psi$ (blue triangles) exhibit similar strong suppression, the $\phi$ meson (red circles) is notably less suppressed at intermediate $p_T$. This species-dependent hierarchy is further emphasized in the right panel for $2.5 < p_T < 5.0$~GeV/c. Across mid-to-high $\langle N_{part} \rangle$, the $\phi$ meson $R_{AA}$ remains near unity or higher, whereas the $\omega+\rho$ and $J/\psi$ follow a significantly more suppressed trend. As detailed in Ref.~\cite{PRC112_2025}, these results suggest that the $\phi$ meson exhibits distinct medium interactions compared to $\omega+\rho$ and $J/\psi$ mesons, highlighting the role of flavor and production mechanisms in shaping forward-rapidity suppression.

The relationship between soft and hard processes is further explored through the multiplicity dependence of $J/\psi$ production in $p+Au$ collisions.
\begin{figure}[h]
\centering
\includegraphics[width=0.48\linewidth]{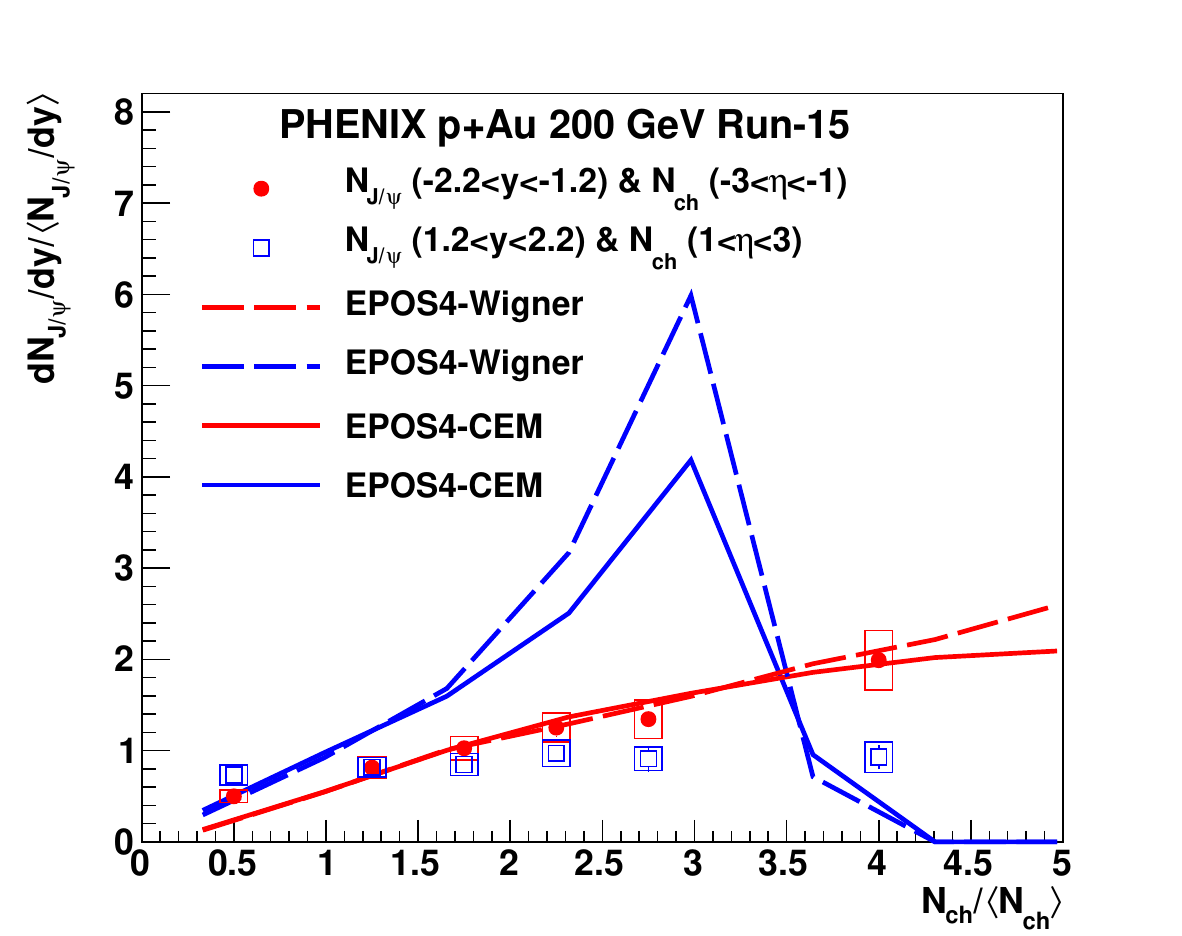}
\includegraphics[width=0.48\linewidth]{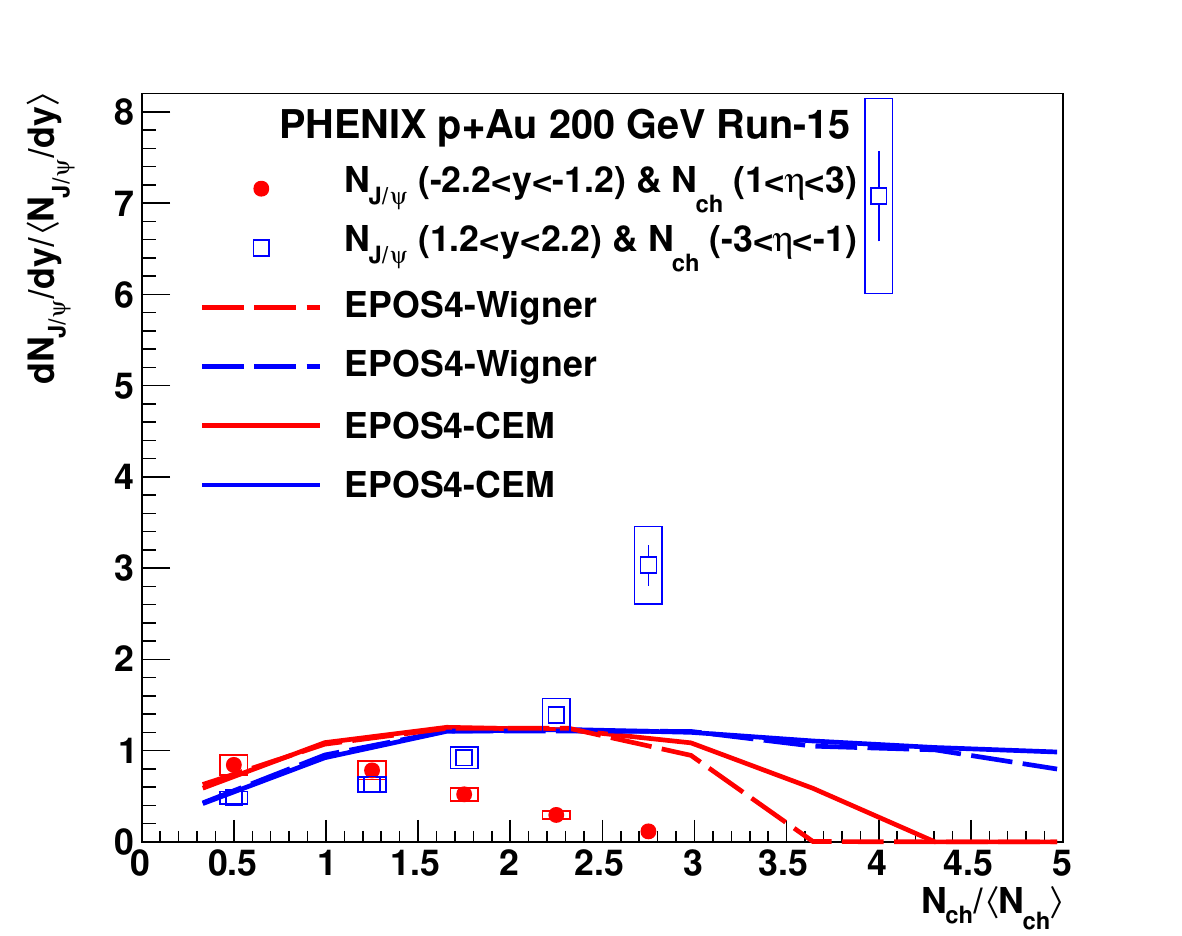}
\caption{Self-normalized $J/\psi$ yields as a function of self-normalized charged-particle multiplicity in $p+Au$ collisions at $\sqrt{s_{NN}} = 200$ GeV. The left and right panels represent different rapidity gap configurations between the $J/\psi$ and the recorded multiplicity. Data points are compared to EPOS4-Wigner and EPOS4-CEM model calculations.}
\label{fig:jpsi_corr}
\end{figure}
As shown in Fig.~\ref{fig:jpsi_corr}, a significant correlation is observed between the self-normalized $J/\psi$ yield ($dN_{J/\psi}/dy / \langle dN_{J/\psi}/dy \rangle$) and the self-normalized charged-particle multiplicity ($N_{ch}/\langle N_{ch}\rangle$). This enhancement is particularly pronounced in the large rapidity gap configuration, suggesting that $J/\psi$ production scales with the global event activity and the number of multiple partonic interactions (MPI).
While the EPOS4-Wigner and EPOS4-CEM models~\cite{ZhaoPrivate,hjb8-pv7p} qualitatively describe the data when both observables are measured in the same rapidity region (Au-going direction), they significantly underpredict the strength of the correlation observed across a large rapidity gap—specifically when the $J/\psi$ is in the $p$-going direction and multiplicity is measured in the Au-going direction. 
This robust correlation over large rapidity intervals suggests that $J/\psi$ production in small systems is strongly influenced by global event characteristics and multiple partonic interactions (MPI), indicating that further refinement of the longitudinal dynamics in current theoretical frameworks is required.

\section{Conclusions and Outlook}
The comprehensive dataset collected by PHENIX provides a broad characterization of particle production across collision systems. The observed scaling with system size and the species dependence at intermediate $p_T$ support a picture in which final-state effects, including partonic energy loss and recombination, play a central role in shaping light hadron production.

The approximate universality of suppression at high $p_T$ across light-flavor species is consistent with expectations from partonic energy loss in the perturbative regime. Meanwhile, the multiplicity dependence of $J/\psi$ production in small systems reveals a strong correlation with global event activity, pointing to an important role of multiple partonic interactions, with contributions from both initial- and final-state effects still to be disentangled.

These results provide important constraints for theoretical models and establish a baseline for future measurements at sPHENIX and the Electron-Ion Collider.

\section*{References}
\bibliography{moriond}

@article{Adcox:2004mh,
  author = {Adcox, K. and others},
  collaboration = {PHENIX},
  journal = {Nucl. Phys. A},
  volume = {757},
  pages = {184},
  year = {2005}
}

@article{Heinz:2013th,
  author = {Heinz, U. and Snellings, R.},
  journal = {Ann. Rev. Nucl. Part. Sci.},
  volume = {63},
  pages = {123},
  year = {2013}
}

@article{Andronic:2017pug,
  author = {Andronic, A. and others},
  journal = {Nature},
  volume = {561},
  pages = {321},
  year = {2018}
}

@article{Adam:2016dau,
  author = {Adam, J. and others},
  collaboration = {ALICE},
  journal = {Nature Phys.},
  volume = {13},
  pages = {535},
  year = {2017}
}

@article{Fries:2003vb,
  author = {Fries, R. J. and others},
  journal = {Phys. Rev. C},
  volume = {68},
  pages = {044902},
  year = {2003}
}

@article{PRC109_2024,
  author = {N. J. Abdulameer and  others},
  journal = {Phys. Rev. C},
  volume = {109},
  pages = {054910},
  year = {2024}
}

@article{PRC112_2025,
  journal = {Phys. Rev. C},
  volume = {112},
  pages = {064918},
  year = {2025},
  author = {N. J. Abdulameer and others}
}

@misc{ZhaoPrivate,
  author = {Zhao, Jiaxing},
  note = {Private communication},
  year = {2025}
}

@article{hjb8-pv7p,
  author = {Werner, K.},
  journal = {Phys. Rev. C},
  volume = {113},
  issue = {1},
  pages = {014911},
  numpages = {13},
  year = {2026},
  publisher = {American Physical Society},
  doi = {10.1103/hjb8-pv7p},
  url = {https://link.aps.org/doi/10.1103/hjb8-pv7p}
}

@article{ADCOX2003469,
journal = {Nucl. Instrum.
Methods Phys. Res., Sec. A},
volume = {499},
number = {2},
pages = {469-479},
year = {2003},
issn = {0168-9002},
doi = {https://doi.org/10.1016/S0168-9002(02)01950-2},
url = {https://www.sciencedirect.com/science/article/pii/S0168900202019502},
author = {K. Adcox and others}
}

@article{PhysRevC.107.014907,
  author = {Abdulameer, N. J. and others},
  collaboration = {PHENIX Collaboration},
  journal = {Phys. Rev. C},
  volume = {107},
  issue = {1},
  pages = {014907},
  numpages = {16},
  year = {2023},
  publisher = {American Physical Society},
  doi = {10.1103/PhysRevC.107.014907},
  url = {https://link.aps.org/doi/10.1103/PhysRevC.107.014907}
}

%%% manually generated bibliography
%\begin{thebibliography}{99}
%\bibitem{ja}C Jarlskog in {\em CP Violation}, ed. C Jarlskog
%(World Scientific, Singapore, 1988).
%\bibitem{ma}L. Maiani, \Journal{\PLB}{62}{183}{1976}.
%\bibitem{bu}J.D. Bjorken and I. Dunietz, \Journal{\PRD}{36}{2109}{1987}.
%\bibitem{bd}C.D. Buchanan {\it et al}, \Journal{\PRD}{45}{4088}{1992}.
%\end{thebibliography}

\end{document}